\documentclass{aip-cp}

\usepackage[numbers]{natbib}
\usepackage{rotating}
\usepackage{graphicx}

\newcommand{\beq}{\begin{equation}} 
\newcommand{\eeq}{\end{equation}} 
\newcommand{\bea}{\begin{eqnarray}} 
\newcommand{\eea}{\end{eqnarray}}

\begin{document}

% Title portion
\title{Discovering New Physics with Voronoi Tessellations}

\author[aff1]{Dipsikha Debnath\corref{cor1}}
%\eaddress[url]{http://www.aip.org}
\author[aff2]{James~S.~Gainer}
\author[aff1]{Doojin~Kim}
\author[aff1]{Konstantin~T.~Matchev}
%\eaddress{anotherauthor@thisaddress.yyy}
\affil[aff1]{Physics Department,
  University of Florida, Gainesville, FL 32611, USA}
\affil[aff2]{Dept. of Physics and Astronomy, University of Hawaii,
  Honolulu, HI 96822, USA}

%\affil[aff1]{Replace this text with an author's affiliation (use complete addresses). Note the use of superscript ``a)'' to indicate the author's e-mail address below. Use b), c), etc. to indicate e-mail addresses for more than 1 author.}
%\affil[aff2]{Additional affiliations should be indicated by superscript numbers 2, 3, etc. as shown above.}
%\affil[aff3]{You would list an author's second affiliation here.}
\corresp[cor1]{dipsikha.debnath@gmail.com}

\maketitle

\begin{abstract}
High energy experimental data can be viewed as a sampling of the relevant phase space.
We point out that one can apply Voronoi tessellations in order to understand the underlying 
probability distributions in this phase space. Interesting features in the data can then
be discovered by studying the properties of the ensemble of Voronoi cells.
For illustration, we demonstrate the detection of kinematic ``edges" in two dimensions,
which may signal physics beyond the standard model. We motivate the algorithm with
some analytical results derived for perfect lattices, and show that the method is further
improved with the addition of a few Voronoi relaxation steps via Lloyd's method. 
\end{abstract}

% Head 1
\section{INTRODUCTION}
In high energy physics, the data is a collection of ``events", which are distributed in phase space, 
${\cal P}$, according to the differential cross-section
\beq
\frac{d\sigma}{d\vec{\bf x}} \equiv f(\vec{\bf x}, \left\{\alpha\right\}).
\label{fdef}
\eeq
Here $\vec{\bf x}\in {\cal P}$ is a phase space point, which is often parameterized in terms of
the momentum components of the final state particles. The set $\left\{\alpha\right\}$ 
is a set of model parameters, e.g., particle masses, widths, couplings, etc. 
The function (\ref{fdef}) consists of two contributions:
\beq
f(\vec{\bf x}, \left\{\alpha\right\}) \equiv f_{SM} (\vec{\bf x}, \left\{\alpha_{SM}\right\})
+f_{NP}(\vec{\bf x}, \left\{\alpha_{NP}\right\}),
\label{f}
\eeq 
where $f_{SM}$ represents the distribution expected from Standard Model (SM) 
processes, a.k.a.~``the background", while $f_{NP}$ is the contribution due to new physics, i.e., ``the signal".
A promising way to look for new physics is to identify structural features in the differential distributions
of the observed events, which might be present in $f_{NP}$, but not in $f_{SM}$.
This idea is similar to the bump-hunting technique in resonance searches, where  we 
look for the Breit-Wigner peak in $f_{NP}$ over the smooth background described by $f_{SM}$.
Even when some of the decay products (e.g., neutrinos or dark matter particles) are invisible in the detector,
one may still look for discontinuities or singularities \cite{Kim:2009si} in the invariant mass distributions of the 
visible particles observed in the detector. Examples of such special features in $f_{NP}$ include:
kinematic endpoints \cite{Hinchliffe:1996iu,Cho:2009ve,Barr:2010zj,Barr:2011xt}, 
kinematic boundaries \cite{Costanzo:2009mq,Burns:2009zi,Matchev:2009iw,Matchev:2009ad,Agrawal:2013uka}, 
kinks \cite{Cho:2007qv,Gripaios:2007is,Barr:2007hy,Cho:2007dh,Burns:2008va} 
and cusps \cite{Han:2009ss,Agashe:2010gt,Han:2012nm,Han:2012nr}
These features are {\em not present} in the background distribution $f_{SM}$.

Here we concentrate on two-dimensional high energy particle physics data, but our
study can be easily generalized to higher dimensions \cite{us}. 
We assume that the signal distribution, $f_{NP}$, changes dramatically or
has a discontinuity in phase space. Such a kinematic boundary or ``edge'' can reveal the 
existence of new particles. Edge detection has been studied in the experimental and 
observational sciences \cite{edgedetection}. However, in particle physics, 
the standard methods of edge detection face several challenges, namely
\begin{enumerate}
\item {\em The data may be sparse.}  Traditional edge detection methods
focus on images, where each pixel contains a data point for a continuous variable (intensity). 
In contrast, in particle physics we look for an edge, which is a possible signature of new physics,
with a comparatively small number of signal events.
\item {\em The analytic form of the distributions $f_{SM}$ and $f_{NP}$ describing the data may be unknown.} 
If the parametric form of the distribution (\ref{f}) is known, we can promptly apply
likelihood methods to determine edges. However, it is usually difficult to get
an exact analytical form for $f_{SM}$, especially in the case of reducible backgrounds,
where detector effects play a major role. Moreover, we cannot be sure, 
{\em \'{a} priori}, that we have correctly assumed the specific new physics model
\cite{Debnath:2014eaa}. Even if we have some idea of where the new physics edges may show up, 
a general procedure is always of greater practical value.
\item {\em The data may be in more than two dimensions.} As we mentioned above, 
edge detection is generally applied to two-dimensional images. However,
in particle physics, multivariate analyses~\cite{multivariate} are present everywhere.
Therefore, in general we will be facing the problem of finding an $(n-1)$-dimensional 
kinematic boundary in an $n$-dimensional parameter space.

\end{enumerate}

Our proposed method for edge detection can handle all three of these
challenges, and may become a useful tool for the experimental
analyses in Run 2 of the CERN Large Hadron Collider (LHC).

\section{A Voronoi Method for Edge Detection}

We start our analysis by making the Voronoi tessellation of some two-dimensional data, 
where each ``event", $i$, represents the corresponding generator point 
for the $i^{\rm{th}}$ Voronoi polygon \cite{voronoi, dirichlet, VT}. This particular 
method of tessellation divides a given volume containing data points
 $\{d_i\}$ into several regions, $\mathcal{R}_i$, such that each $\mathcal{R}_i$ 
contains exactly one data point, $d_i$, and for any point
 $p \in \mathcal{R}_i$, $d_i$ is the nearest data point. 

We focus to identify edge features such as discontinuities \cite{1d} without assuming the 
exact knowledge of the $f_{NP}$ distribution. There exist several edge detection algorithms 
for binned data \cite{canny}. Our Voronoi method of edge detection avoids binning and 
includes the following steps:
 \begin{enumerate}
\item Construct the Voronoi tessellation for the data set.
\item Compute relevant attributes of the Voronoi cells.
\item (Optionally) use the information from the previous step 
to further process the data in some way.
\item Use some criterion to flag ``candidate'' edge cells.
\item Identify an edge from the collection of edge cell candidates.
\end{enumerate}

\begin{figure}[t] 
%\begin{center}
\includegraphics[height=0.2 \columnwidth]{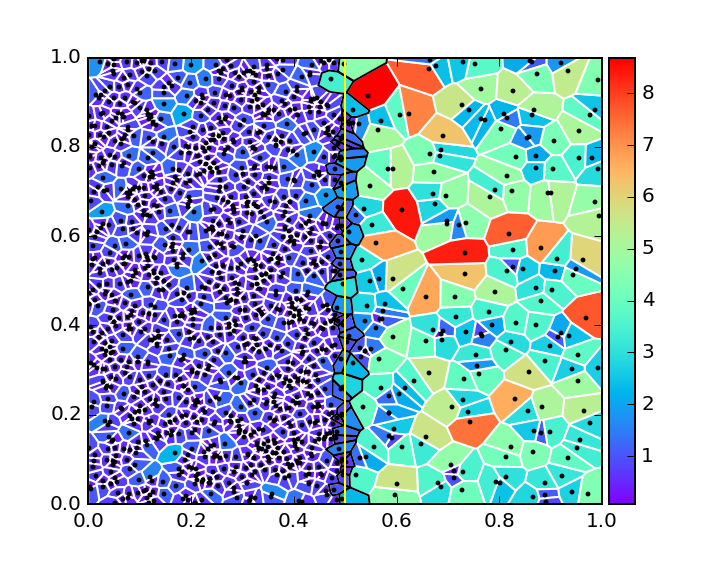} 
\includegraphics[height=0.2 \columnwidth]{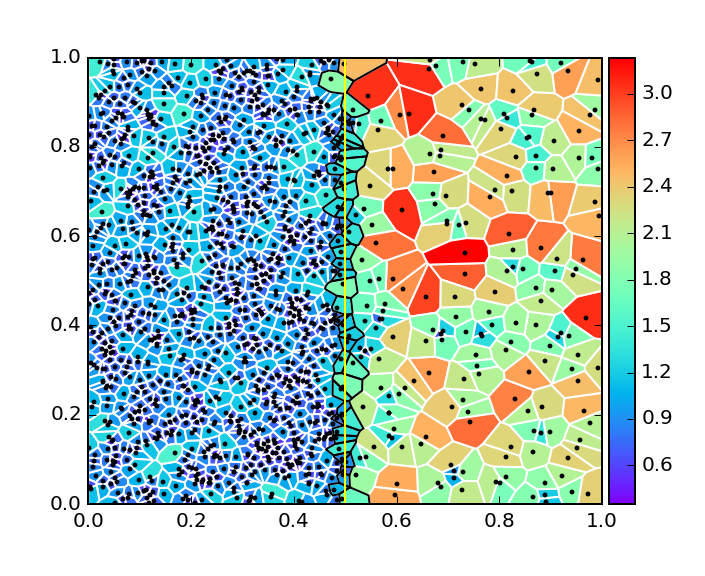}  
\includegraphics[height=0.2 \columnwidth]{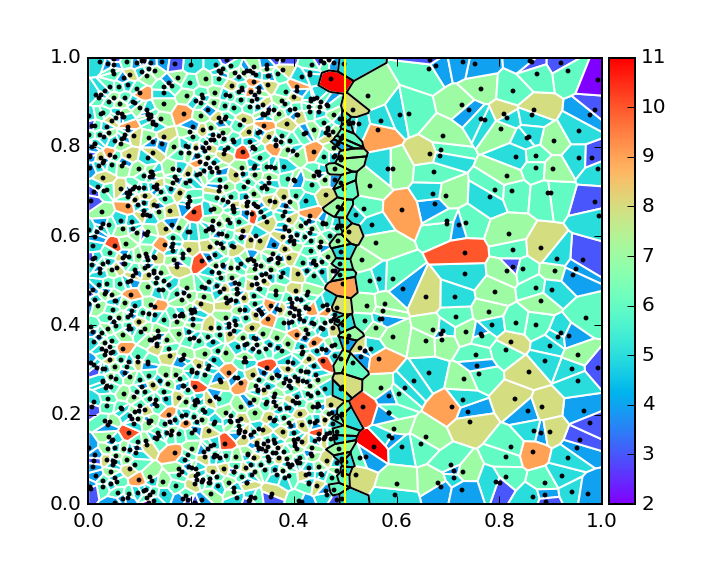} 
\includegraphics[height=0.2 \columnwidth]{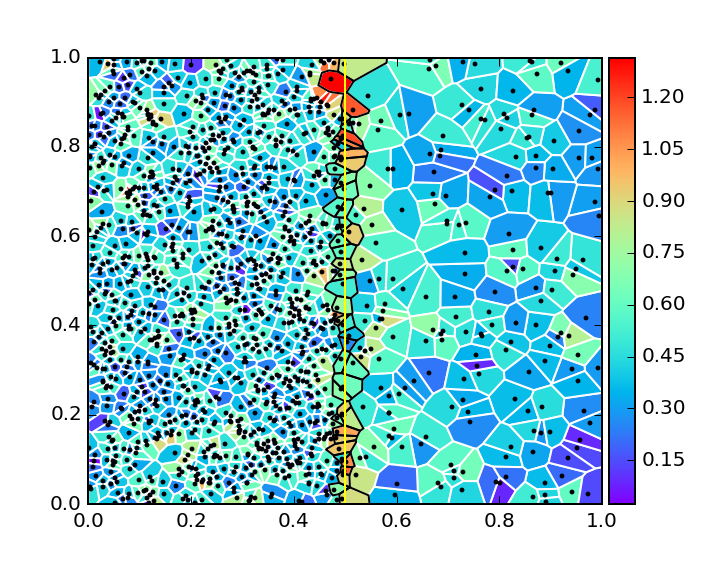}
%\end{center}
\caption{Voronoi tessellations for 1400 data points selected from the 
  probability density (\ref{fstep}) with $\rho=6$. The Voronoi polygons are
  color-coded by their area (left), perimeter (middle left),
  number of neighboring polygons (middle right) or scaled variance
  (\ref{defvar}) (right). 
\label{fig:temp} }
\end{figure}

Some useful intuition can be gained from the following toy example.
We generate 1400 points according to the probability distribution
\beq
f(x,y) =
\frac{2}{1+\rho}
\left[ \rho\, H(0.5-x) + H(x-0.5) \right].
\label{fstep}
\eeq
within the unit square.
In eq.~(\ref{fstep}), $H(x)$ is the Heaviside step function and $\rho$ is a constant density ratio. 
The resulting Voronoi tessellation is shown in Figure~\ref{fig:temp},
where the color-code for each Voronoi polygon represents some standard property,
such as area, perimeter, or number of immediate neighbors. The square is divided into 
left (L) and right (R) regions of constant, but unequal densities. 
Our goal is to spot the the vertical edge at $x=0.5$ (yellow solid line) 
where the density sharply changes from one region to other. 
For convenience, we outline the boundaries of the Voronoi cells, 
crossing the edge at $x=0.5$ as black and the remaining Voronoi cells away from the edge as white.
%%%%%%%%%%%%%%%%%%%%%BEGIN FIGURE%%%%%%%%%%%%%%%%%%%%%%%%%
%%%%%%%%%%%%%%%%%%%%%END FIGURE%%%%%%%%%%%%%%%%%%%%%%%%%  

The two leftmost panels of Figure~\ref{fig:temp} show that the area and perimeter 
of the Voronoi polygons are somewhat correlated, while the middle right panel reveals that
the typical number of nearest neighbors is similar in the two bulk regions. Therefore, these properties or aspects
of Voronoi polygons cannot help in finding the edge cells (outlined in black). 
This is why we introduce a new variable, the scaled standard deviation of the areas of the neighboring cells,
\beq
\frac{\sigma_a}{\bar{a}} \equiv \frac{1}{\bar{a}}\, \sqrt{\sum_{n\in N_i} \frac{\left(a_n-\bar{a}\right)^2}{|N_i|-1}},
\label{defvar}
\eeq
where $N_i$ is the set of neighbors of the $i$-th Voronoi polygon,
and $\bar{a}(N_i)$ is their mean area. The scaled standard deviation is quite successful
in picking out edge cells and this can be visualized in the rightmost panel in 
Figure~\ref{fig:temp}. Thus we choose (\ref{defvar}) as our 
main selection variable\footnote{This is not the only option, however --- we have
investigated a number of other promising variables which will be discussed
in a longer publication \cite{us}.}.

%%%%%%%%%%%%%%%%%%%%%BEGIN FIGURE%%%%%%%%%%%%%%%%%%%%%%%%%
\begin{figure}[t] 
\includegraphics[height=0.19 \columnwidth]{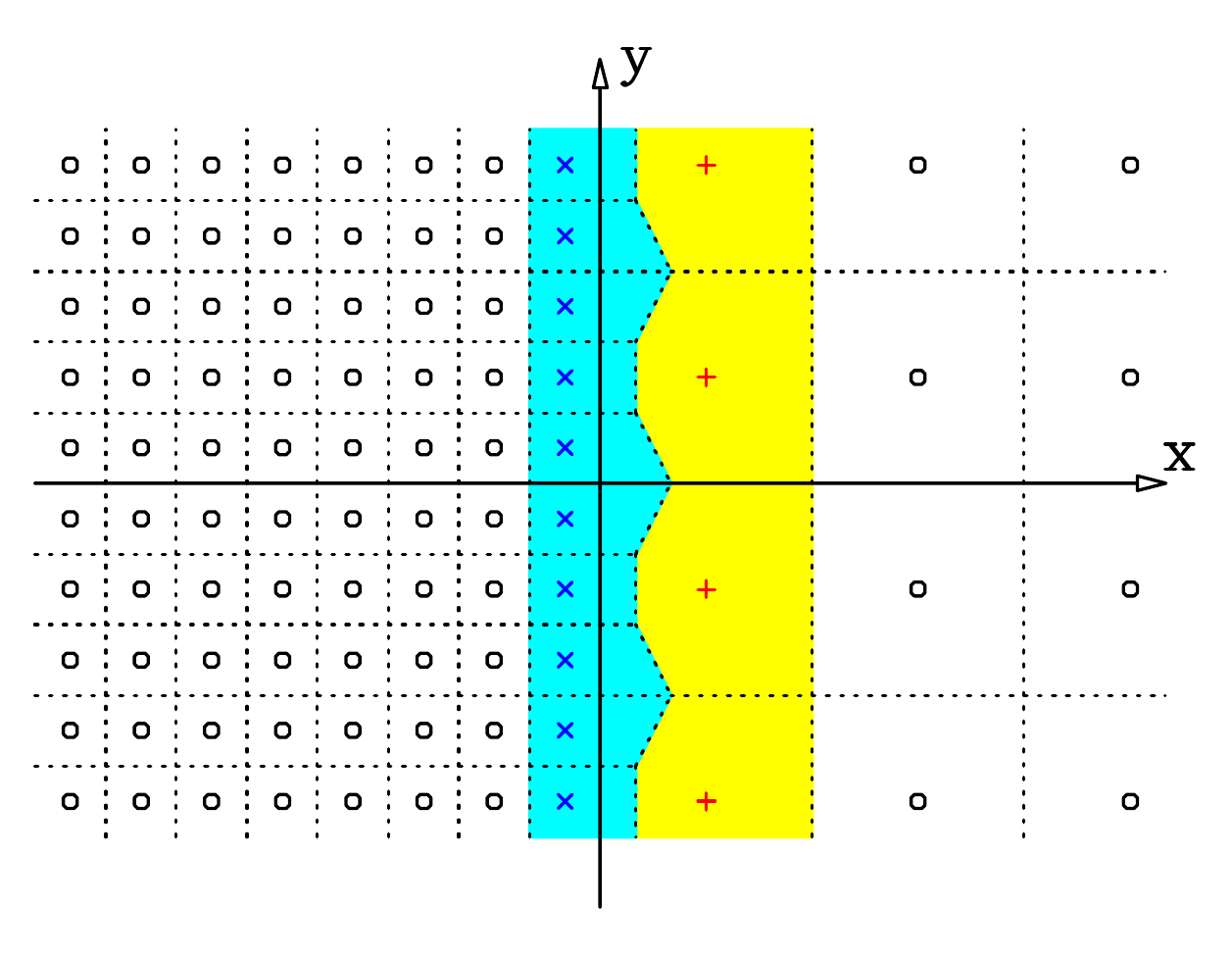}
\includegraphics[height=0.19 \columnwidth]{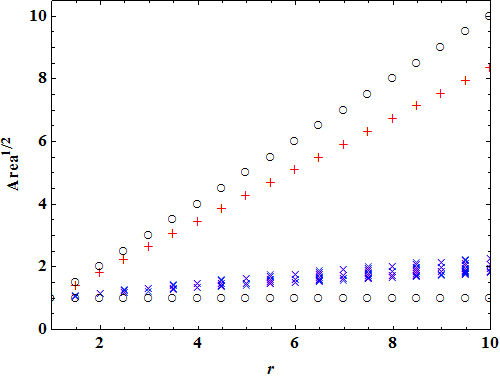} 
\includegraphics[height=0.19 \columnwidth]{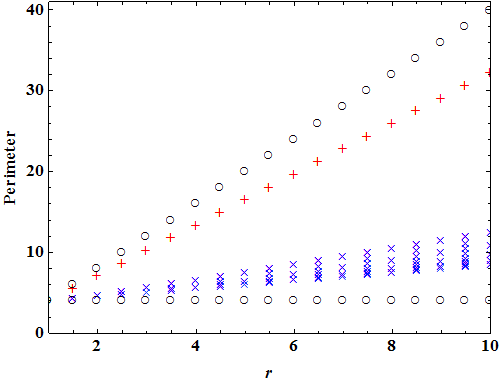}
\includegraphics[height=0.19 \columnwidth]{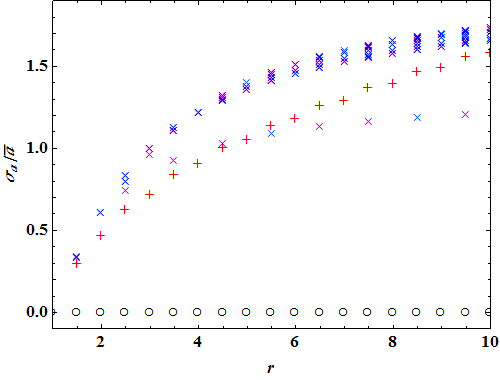} 
\caption{ A regular lattice (\ref{lattice}) generated for linear density ratio
  $r=3$ (left), and the dependence on $r$ of
  several parameters of interest, namely cell area (middle left),
  cell perimeter (middle right) and scaled variance (right).
  Black circles indicate bulk cells, while blue $\times$ (red $+$) symbols
 denotes edge cells in the L (R) region.
\label{fig:reg} }
\end{figure}
%%%%%%%%%%%%%%%%%%%%%END FIGURE%%%%%%%%%%%%%%%%%%%%%%%%%  

In order to understand the above results %generated from the stochastic process
analytically, we consider a perfect grid of points which follows the probability distribution (\ref{fstep}). 
The grid is generated by two integers $n$ and $m$ as
\beq
\vec{R} = \left[ \left( n+0.5 \right) \hat x + \left( m+0.5\right) \hat y\right]
\left[H(-n) + rH(n)\right],
\label{lattice}
\eeq
where the vectors $\hat x$ and $\hat y$ form an orthonormal basis and $r\equiv\sqrt{\rho}$
is the corresponding linear density ratio.
The left panel of Figure~\ref{fig:reg} shows an example grid for $r=3$. 
We highlight the two columns of edge cells: in the L region (blue $\times$ symbols) and the R region (red $+$ symbols).
The other three panels in Figure~\ref{fig:reg} show the behavior of
some of their properties as a function of $r$.  For the case of area and perimeter
we notice that the values for edge cells are intermediate between the two
bulk values. However, the scaled standard deviation is exactly zero for both bulk regions, 
and nonzero for the edge region, thus offering the possibility for good discrimination.

\section{Voronoi relaxation via Lloyd's algorithm}

As we are dealing with a stochastic process, statistical fluctuations are unavoidable in the data.
In particular, in Figure~\ref{fig:temp} we can easily spot a few bulk cells having relatively high 
values of $\sigma_a/\bar{a}$. This is why we introduce the idea of ``smoothing" the data 
by applying a few iterations of Lloyd's algorithm \cite{lloyd}, where at each iteration, 
the generator point is replaced by the centroid of the corresponding Voronoi 
cell.\footnote{An alternative approach, illustrated below in Figure~\ref{fig:susy},
would be to leave the original Voronoi tessellation 
intact, but extend the calculation of (\ref{defvar}) to include next-to-nearest neighbors, next-to-next-to-nearest neighbors, etc.}
%%%%%%%%%%%%%%%%%%%%%BEGIN FIGURE%%%%%%%%%%%%%%%%%%%%%%%%%
\begin{figure}[ht] 
\includegraphics[height=0.24 \columnwidth]{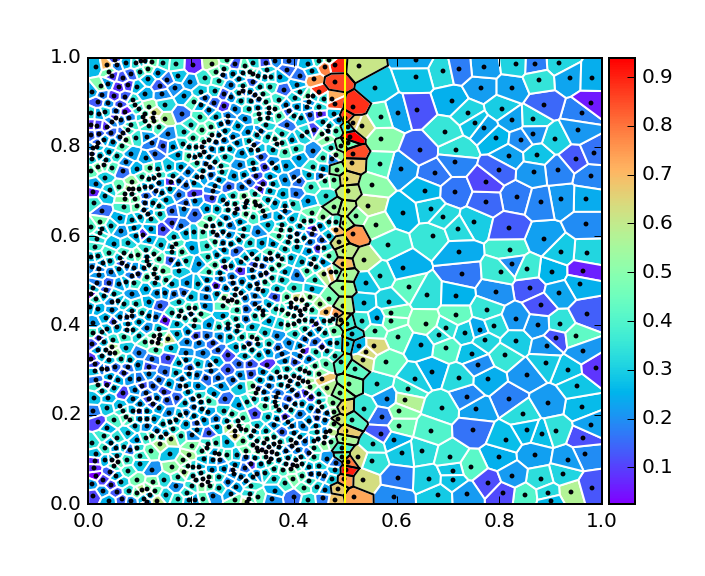} 
\includegraphics[height=0.24 \columnwidth]{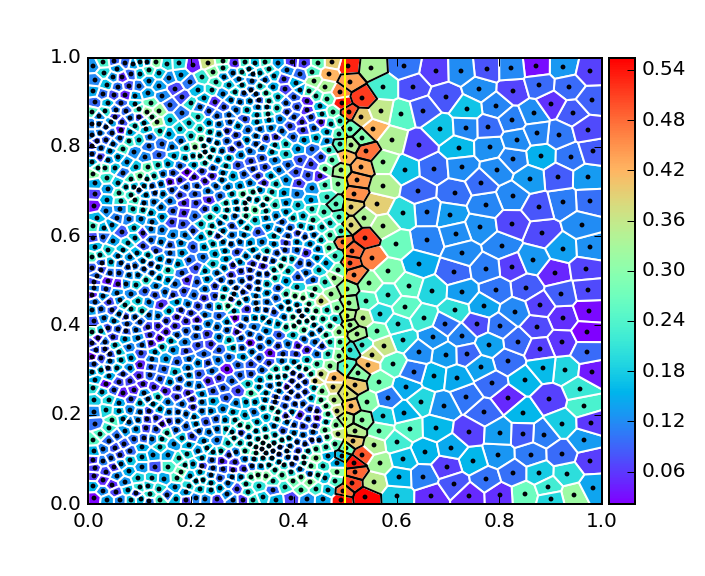}
\caption{The evolution of the Voronoi tessellation shown in
  Figure~\ref{fig:temp}
  after one (left panel) and five (right panel) applications of Lloyd's
  algorithm.
  The cells are color-coded by the scaled variance (\ref{defvar}). 
\label{fig:movie} }
\end{figure}
%%%%%%%%%%%%%%%%%%%%%END FIGURE%%%%%%%%%%%%%%%%%%%%%%%%%  
Figure~\ref{fig:movie} shows the Voronoi tessellation after one (left panel) and five (right panel) Lloyd iterations.
We find that the Voronoi polygons become more regularly shaped after relaxation and the fluctuations on each side of the 
boundary are washed out. Most importantly, the values of the scaled standard deviation (\ref{defvar}) for the edge cells are enhanced
relative to the rest.

Figure~\ref{fig:movie} also shows that as a result of the Voronoi relaxation,
the data points from the dense L region flow towards the
relatively sparse R region. Consequently, the edge cells with high $\sigma_a/\bar{a}$
are displaced from their original locations (near the vertical yellow line).
For this reason, once we select edge cell candidates after a certain number 
of Lloyd iterations, we need to trace them back to their original locations before
doing any further quantitative data analysis.

%%%%%%%%%%%%%%%%%%%%%BEGIN FIGURE%%%%%%%%%%%%%%%%%%%%%%%%%
\begin{figure}[ht] 
\includegraphics[height=0.24 \columnwidth]{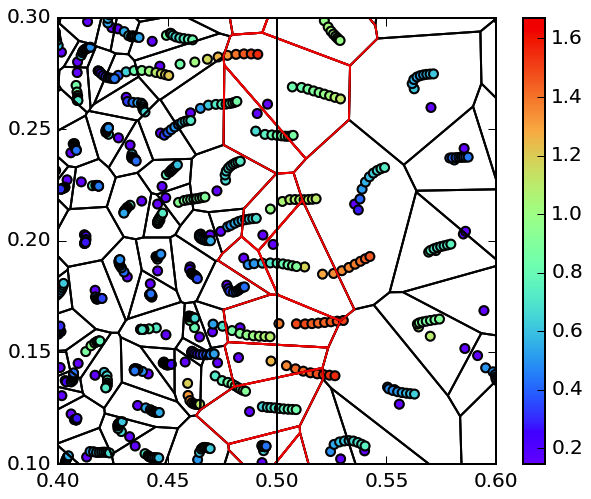} 
\includegraphics[height=0.24 \columnwidth]{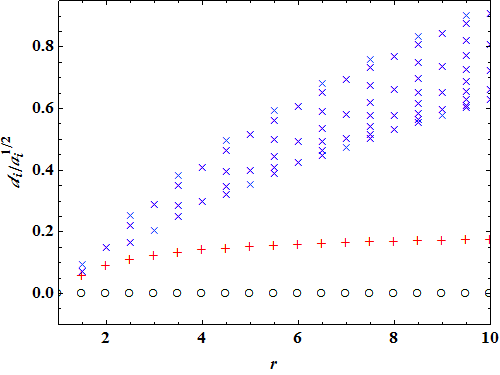}
\caption{ Left: a zoomed-in region near the vertical edge, which shows the
  originally generated points and their subsequent locations after repeated
  application of Lloyd's algorithm.
The points are color-coded by scaled displacement, $d_i/\sqrt{a_i}$.
Right: Predictions for $d_i/\sqrt{a_i}$ after applying Lloyd's algorithm once,
as a function of $r$, for
the case of a regular lattice (\ref{lattice}).
\label{fig:worms} }
\end{figure}
%%%%%%%%%%%%%%%%%%%%%END FIGURE%%%%%%%%%%%%%%%%%%%%%%%%% 

By comparing the displacements $d_i$ of the generator points, 
we notice that the edge points tend to be displaced the farthest. 
We can use this as an alternative tagging method. To quantify this criterion, 
we define a dimensionless variable, the scaled displacement,
$d_i/\sqrt{a_i}$, where we normalize by the square root of the cell area, $a_i$. 
The left panel in Figure~\ref{fig:worms} gives a closer view of one 
representative area near the edge and shows the result of several successive Lloyd iterations.
The color code indicates that the scaled displacement is indeed a useful quantity,
just like the scaled standard deviation (\ref{defvar}). 
We confirm this by showing in the right panel of Figure~\ref{fig:worms} the exact result for the perfect grid (\ref{lattice}).

\begin{figure}[ht] 
\includegraphics[height=0.24 \columnwidth]{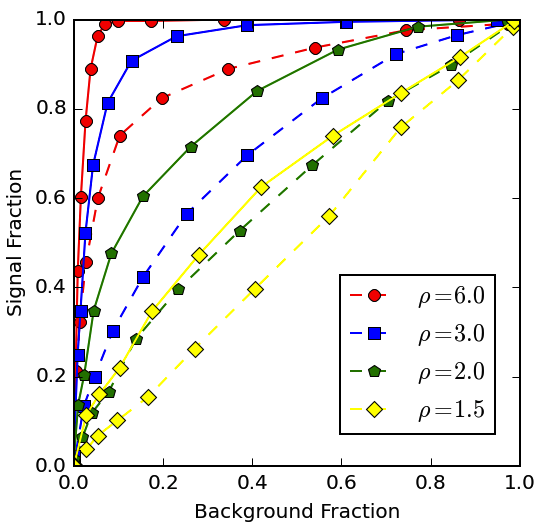} 
\includegraphics[height=0.24 \columnwidth]{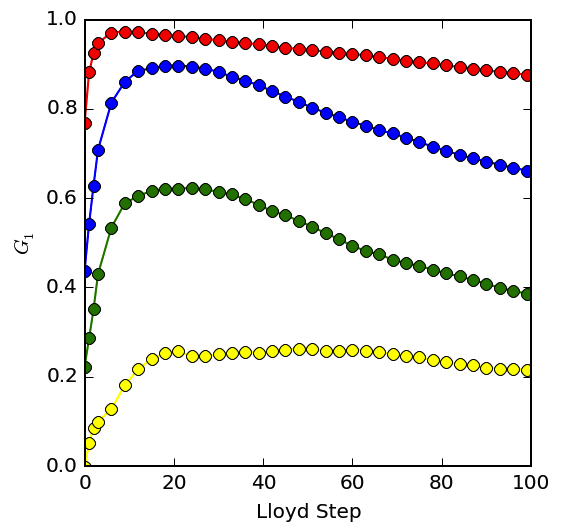}
\caption{ Left: ROC curves $\varepsilon_S(\varepsilon_B)$ obtained using
  (\ref{defvar}) as the discriminating variable.
  Right: The Gini index (\ref{Gini}) found from the ROC curve obatined after
  the given number of Lloyd iterations.
\label{fig:ROC} }
\end{figure}

We study the efficiency of our edge detection algorithm by analyzing  ROC curves \cite{ROC}. We generate a large
dataset for (\ref{fstep}), where we consider the edge cells as ``signal" and the bulk cells as ``background".
We plot the signal selection efficiency, $\varepsilon_S$, versus the background efficiency,
 $\varepsilon_B$, for different values of the minimum cut on the variable (\ref{defvar}).
Several $\varepsilon_S(\varepsilon_B)$ curves, for different values of the density ratio $\rho$, 
and either with (solid) or without (dashed) Lloyd relaxation are shown in Figure~\ref{fig:ROC}. 
The ROC curves reveal that the algorithm is more efficient for higher density contrasts between the two regions. 
In addition, the Voronoi relaxation leads to a significant improvement of the result.

%%%%%%%%%%%%%%%%%%%%%BEGIN FIGURE%%%%%%%%%%%%%%%%%%%%%%%%%
%%%%%%%%%%%%%%%%%%%%%END FIGURE%%%%%%%%%%%%%%%%%%%%%%%%%  

The accuracy of our selection criteria is quantified by using the standard area under the curve 
\cite{AUROC} (AUROC) as represented by the Gini coefficient
\beq
G_1\equiv 2\, {\rm AUROC} - 1 = 2\int_0^1 d\varepsilon_B \times \varepsilon_S(\varepsilon_B) -1,
\label{Gini}
\eeq
where a value of $1$ is obtained from the ROC curve of a perfectly discriminating variable, while
a value of $0$ corresponds to a totally random selection of events. The right panel of Figure~\ref{fig:ROC}
shows the dependence of $G_1$ on the number of Lloyd steps. We see that the sensitivity
improves dramatically within the first few iterations, and reaches an optimum plateau, after which the
power of the test is degraded as the Voronoi grid begins to asymptote to the centroidal tessellation.

\section{An example from supersymmetry} 

We apply our proposed edge detection method to a standard benchmark example from supersymmetry; 
squark pair production at the 13 TeV LHC. We consider events where one squark undergoes 
a long cascade decay through a heavy neutralino, $\tilde \chi^0_2$; a slepton, $\tilde \ell$; and a light neutralino, $\tilde\chi^0_1$;
while the other decays directly to $\tilde\chi^0_1$. The mass spectrum is chosen to be 
$m_{\tilde q}=400$ GeV, $m_{\tilde\chi^0_2}=300$ GeV, 
$m_{\tilde \ell}=280$ GeV, and $m_{\tilde\chi^0_1}=200$ GeV.
The invariant mass distributions of the final state particles, the two jets and
the two leptons, exhibit kinematic edges.
%%%%%%%%%%%%%%%%%%%%%BEGIN FIGURE%%%%%%%%%%%%%%%%%%%%%%%%%
\begin{figure}[t]
\includegraphics[height=3.3cm]{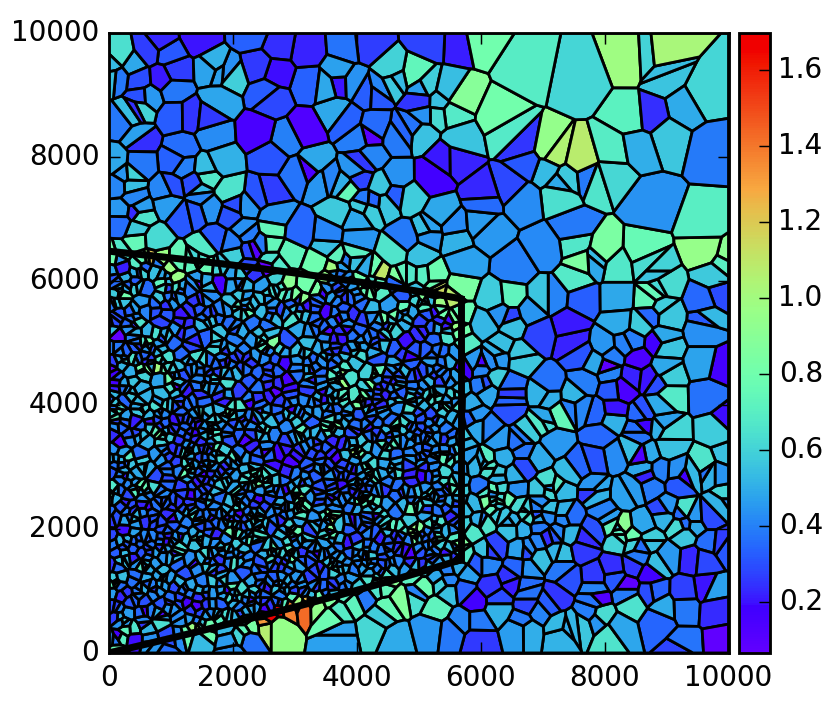}
\includegraphics[height=3.3cm]{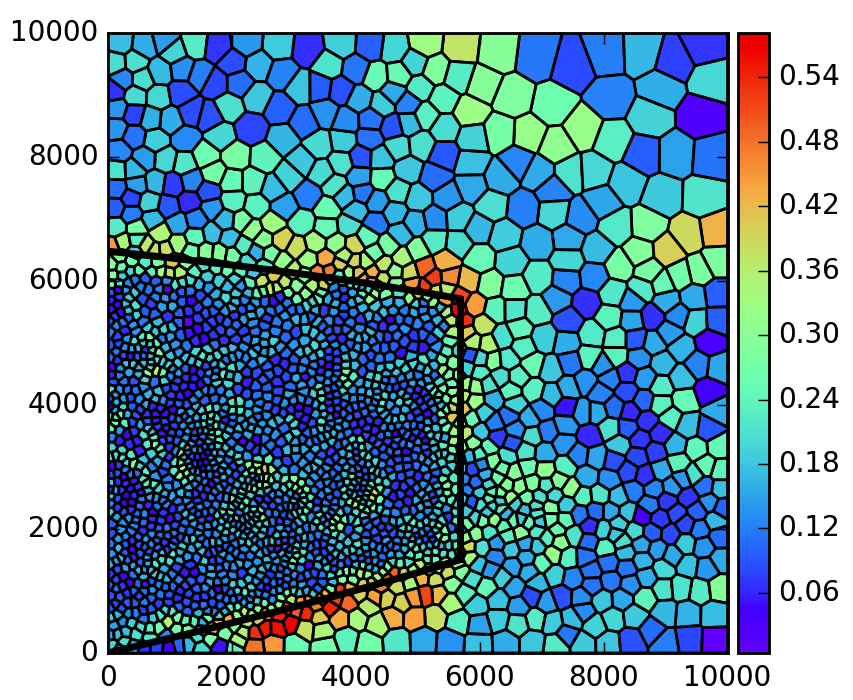}
\includegraphics[height=3.3cm]{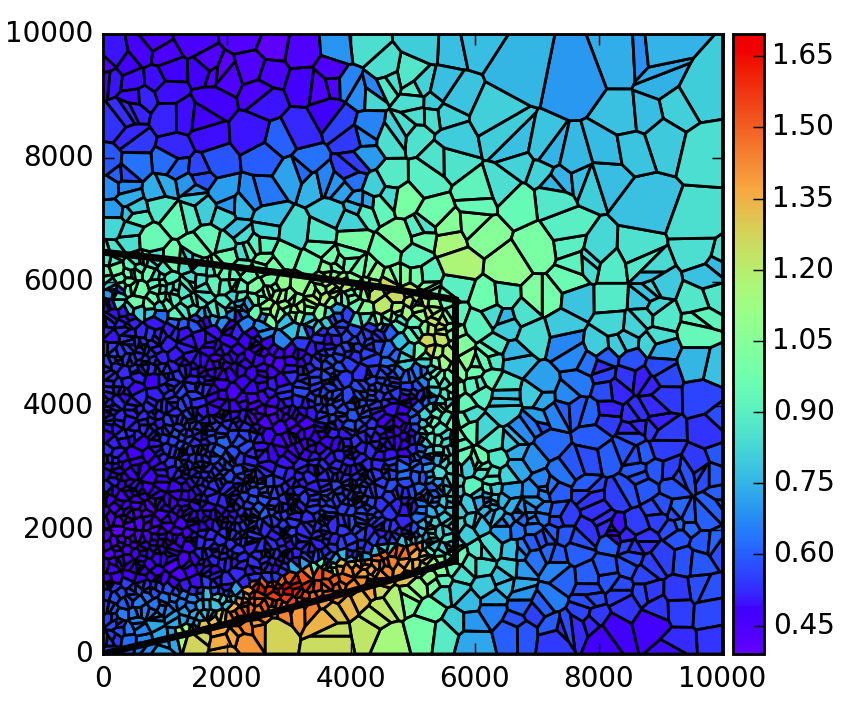}
\includegraphics[height=3.3cm]{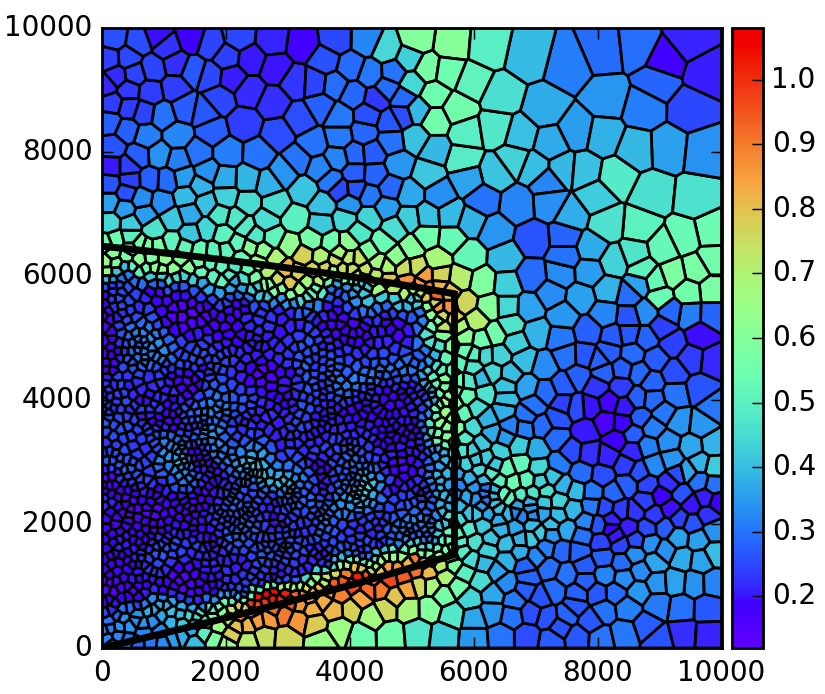}
\caption{ Voronoi tessellations for the supersymmetry example described
  in the text.
%in the $(m^2_{\ell\ell}, (m^2_{j\ell\ell}-m^2_{\ell\ell})/6)$ plane.
\label{fig:susy} }
\end{figure}
%%%%%%%%%%%%%%%%%%%%%END FIGURE%%%%%%%%%%%%%%%%%%%%%%%%%  

In particular, here we consider the dilepton invariant mass, $m_{\ell\ell}$,
and the three-body jet-lepton-lepton invariant mass, $m_{j\ell\ell}$.
In Figure~\ref{fig:susy} we use the $(m^2_{\ell\ell}, (m^2_{j\ell\ell}-m^2_{\ell\ell})/6)$ plane
for plotting convenience. The solid black line in Figure~\ref{fig:susy} \cite{Matchev:2009iw, Lester:2006cf} 
marks the location of the kinematic endpoint for signal events with the correct jet assignment. 
(The lack of knowledge of the charge of the jet creates a two-fold combinatorial ambiguity. Thus, for each event there are two 
entries in the plot.) The main SM background from $t\bar{t}$ dilepton events is also included here.

In Figure~\ref{fig:susy}, the Voronoi cells are color coded by their scaled standard deviation (\ref{defvar}). 
In the left panel we exhibit the original data, while in the middle left panel we show the data after 5 Lloyd iterations. 
We reconsider the original data and extend the calculation of (\ref{defvar}) including up to 5 tiers of nearest neighbors,
showing the resulting plot in the middle right panel. We observe that either Voronoi relaxation
or the addition of more tiers of neighboring cells reduces the fluctuation and sharpens the edge.
Finally, in the rightmost panel of Figure~\ref{fig:susy} we show the result after 3 Lloyd iterations
{\em and} also including 3 tiers of neighbors in the calculation of (\ref{defvar}).

\section{Summary} We argue that the discovery of new kinematic features is 
an essential step in the discovery of physics beyond the standard model at the LHC and
advocate the use of Voronoi methods for this purpose.
The great flexibility of Voronoi methods is a blessing for the experimentalist; 
many useful properties of the Voronoi cells can be used to construct powerful variables
tailored to specific new physics scenarios. A voluminous, quantitative study of the
many options available to the experimenter will be presented in a companion paper~\cite{us}.

\section{ Acknowledgements} We thank S.~Das, C.~Kilic, Z.~Liu, R.~Lu, P.~Ramond, X.~Tata, 
J.~Thaler, B.~Tweedie, and D.~Yaylali for useful discussions.
Work supported in part by U.S. Department of Energy, in part by Grant
DE-SC0010296.  DK acknowledges support by LHC-TI postdoctoral fellowship 
under grant NSF-PHY-0969510.

% References

%\nocite{*}
%\bibliographystyle{aipnum-cp}%
%\bibliography{sample}%

\end{document}